\documentclass[aps,twocolumn,showkeys,nofootinbib]{revtex4}

\usepackage{graphics}
\usepackage{bm}
\usepackage{amssymb}

\begin{document}

\title{Infinitesimal propagation equation for atmospheric decoherence with multiphoton correlations}
\author{Filippus S. Roux}
\affiliation{CSIR National Laser Centre, P.O. Box 395, Pretoria 0001, South Africa}
\email{fsroux@csir.co.za}

\begin{abstract}
Previously a set of coupled first order differential equations were derived for the decoherence of a pair of spatial mode entangled photons, propagating along different paths through turbulence. Here we extend this analysis to the situation where both photons travel along the same path, which introduces the effect of multiple photon correlations. The resulting equation now contains additional terms that take these multiphoton correlations into account. At the same time, we provide a more thorough formulation of the quantized field, starting from a Lorentz invariant formulation, which is then explicitly broken by the choice of a particular propagation direction. The effect of the latter improvement in the quantization on the form of the final equation is minimal.
\end{abstract}

\maketitle

\section{Introduction}

While the orbital angular momentum (OAM) states of a photon allow higher dimensional representations of quantum information \cite{zeil1,qkdn,torres}, an entangled photonic state that is encoded in terms of OAM states will suffer degradation of the encoded information when it propagates through a turbulent medium as a result of the decoherence of the entanglement that is caused by the scintillation. To design a quantum communication system one therefore needs to understand the decoherence process of spatial mode entanglement in a turbulent atmosphere.

Recently a theoretical framework, in the form of an infinitesimal propagation equation (IPE), has been developed to handle this problem \cite{ipe}. Although it extended previous work \cite{turboam1,qturb1,qturb4,qturb3}, which modeled the turbulence with a single phase function, the framework in \cite{ipe} considered only single photons going through a particular turbulent atmosphere at a time. The effect of multiple photons going through the same medium has therefore been neglected. Here we extend the framework to include the additional correlations that are introduced when multiple photons propagate through the same medium. As a result we obtain additional terms in the final expression for the IPE. These terms serve to link different photons with each other. 

At the same time we also provide a more thorough discussion of the quantization of the electric field. Previously \cite{ipe}, the quantization of the electric field was expressed in the popular way \cite{mw}. However, it can be shown that the quantum states that are defined in this way are manifestly not Lorentz covariant \cite{ps} (under a Lorentz boost the orthogonality relation for these quantum states transforms into an expression that is different from the original expression).

On the other hand, one can argue that the final expression would inevitable be non-covariant as a result of the explicit violation of Lorentz covariant when one either fixes the propagation direction or enforces the field to be monochromatic. The question then is whether the final expression could be derived consistently from an initial Lorentz covariant formulation. In other words, does the final expression look the same when one starts with the usual non-covariant quantization as it would when one starts from a Lorentz covariant quantization and then goes, in a consistent manner, through the steps that explicitly breaks the Lorentz covariance? We find that there is a minor difference between these expressions (the integrals in the latter case contain an additional factor of $1/k_z$), but in the paraxial limit this difference becomes insignificant.

\section{Field quantization}
\label{fqua}

In general, quantum basis states are defined in the Fourier domain where they represent a plane wave basis and depend on the three components of the propagation vector $k_x$, $k_y$ and $k_z$ and on the angular frequency $\omega$. Due to the vacuum dispersion relation $\omega = c|{\bf k}|$, any three of the four quantities $\omega$, $k_x$, $k_y$ and $k_z$ will fix the fourth. Therefore, the phase space integrals for propagating (on shell) fields only run over three of these quantities. It is natural to select $k_x$, $k_y$ and $k_z$ as the three integration variables, which then fixes the value of $\omega$. In this way one can define all the quantities in an explicitly Lorentz covariant manner. We'll start with the expressions in this form. 

In the current scenario where we consider the decoherence of OAM entangled states in atmospheric turbulence, we find that the system is not one that evolves in time, but rather in space \cite{ipe}. The quantum states represent excitations of an optical beam propagating in a particular direction through space, which we define as the $z$-axis. In this situation it is more convenient to integrate over $\omega$, $k_x$ and $k_y$ and thereby fix $k_z$. The reason is that the system's evolution is considered as a function of the propagation distance and not as a function of time. Moreover, the input is specified as a function on a two-dimensional plane perpendicular to the propagation direction for all time (as opposed to a three-dimensional function through all space at a fix initial point in time). Such an input can also be expressed in the Fourier domain as a function of $k_x$, $k_y$ and $\omega$. In the monochromatic case one can fix $\omega$ and then end up with a two-dimensional function of $k_x$ and $k_y$.

It should be noted that by fixing the $z$-axis as the propagation direction, one explicitly breaks the rotation symmetry, which forms part of the Lorentz symmetry. As a result our final expression implies an explicit breaking of Lorentz invariance. Nevertheless, our initial formulation should still be Lorentz invariant to ensure consistency. We therefore start with expressions that are defined in terms of Lorentz covariant integrals over $k_x$, $k_y$ and $k_z$. Performing a change of variables, we then end up with quantities that are defined in terms of integrals over $\omega$, $k_x$ and $k_y$.

\subsection{Three-dimensional momentum states}

The creation and annihilation operators are defined in terms of how they create or annihilate the momentum states. The Lorentz covariant orthogonality condition for the three-dimensional momentum basis is given by \cite{ps}\footnote{For a boost along, say, the $z$-direction, given by $k_z'=\gamma(k_z+\beta\omega/c)$ and $\omega'=\gamma(\omega+c\beta k_z)$, one finds that the Dirac delta function for the $z$-component transforms as 
\[\delta(k_z-k_{z1})=\delta(k_z'-k_{z1}') {{\rm d}k_z'\over{\rm d} k_z}=\delta(k_z'-k_{z1}'){\omega'\over \omega} , \]
where we made use of the vacuum dispersion relation. To avoid this non-covariant transformation, the expression for the orthogonality condition needs an extra factor of $\omega$.} 
\begin{equation}
\langle {\bf k}_1,r|{\bf k}_2,s\rangle = \omega\ (2\pi)^3 \delta_{r,s}\ \delta({\bf k}_1-{\bf k}_2) ,
\label{inprodk0}
\end{equation}
where ${\bf k}=k_x \hat{x} + k_y \hat{y} + k_z \hat{z}$ and the $r$ and $s$ represent the spin state, which we include here for the sake of completeness. These basis states are generated or destroyed by creation and annihilation operators according to
\begin{eqnarray}
\langle {\bf k},s| & = & \sqrt{\omega}\ \langle 0 | a_s({\bf k}) \nonumber \\ 
| {\bf k}, s \rangle & = & \sqrt{\omega}\ a_s^{\dag}({\bf k}) | 0 \rangle ,
\label{toest}
\end{eqnarray}
so that they obey the commutation relation
\begin{equation}
\left[ a_s({\bf k}_1), a_r^{\dag}({\bf k}_2) \right] = (2\pi)^3 \delta_{s,r}\ \delta( {\bf k}_1- {\bf k}_2 ) .
\label{commut}
\end{equation}

We define an identity operator that is resolved in terms of the three-dimensional momentum basis \cite{ps}
\begin{equation}
1 = \sum_s \int |{\bf k},s\rangle \langle{\bf k},s|\ {{\rm d}^3 k\over\omega\ (2\pi)^3} .
\label{ident}
\end{equation}
where $\omega$ is given by
\begin{equation}
\omega = c \left( k_z^2+k_x^2+k_y^2 \right)^{1/2} .
\label{kzfunk}
\end{equation}
Integral signs without explicit integration boundaries always imply integration from $-\infty$ to $\infty$.

The identity operator defined in Eq.~(\ref{ident}) can be used to find the expansions of arbitrary one-photon states in terms of the three-dimensional momentum states
\begin{eqnarray}
|\Psi \rangle & = & \sum_s \int |{\bf k},s\rangle \Psi({\bf k},s)\ {{\rm d}^3 k\over\omega\ (2\pi)^3}  \label{bra3d} \\
\langle \Psi | & = & \sum_s \int \Psi^*({\bf k},s) \langle {\bf k},s |\ {{\rm d}^3 k\over\omega\ (2\pi)^3} , \label{ket3d}
\end{eqnarray}
where the momentum space wave function is given by 
\begin{equation}
\Psi({\bf k},s) = \langle {\bf k},s |\Psi \rangle 
\label{binnep}
\end{equation}
and $^*$ represents the complex conjugate.

The one-photon states are normalized so that $\langle \Psi |\Psi \rangle=1$, which then implies that
\begin{equation}
1 = \sum_s \int \left| \Psi({\bf k},s) \right|^2\ {{\rm d}^3 k\over\omega\ (2\pi)^3} .
\label{norm0}
\end{equation}

\subsection{Choosing a propagation direction}

Now we fix the propagation direction to be the $z$-direction and redefine the quantities in terms of the two-dimensional momentum-energy states $|{\bf K},\omega,s\rangle$, instead of the three-dimensional momentum states $|{\bf k},s\rangle$, where ${\bf K}=k_x \hat{x} + k_y \hat{y}$ and $s$ denotes the spin state. By fixing a specific direction for propagation we explicitly break rotation invariance and, by implication, also Lorentz invariance.

From the vacuum dispersion relation $\omega^2 = c^2|{\bf k}|$ it follows that $\omega\ {\rm d}\omega = c^2 k_z\ {\rm d}k_z$ or
\begin{equation}
{\rm d}k_z = {\omega\over c^2 k_z}\ {\rm d}\omega .
\label{wvskz}
\end{equation}

One needs to be careful with the negative frequencies and negative $k_z$'s and how they map to each other. In the applications that we consider, the angular spectrum of the beam only contains nonzero components in the positive $k_z$-direction. Fortunately, since both sides of the $\omega$-axis map into the positive side of the $k_z$-axis, we do not encounter a problem.

Applying the change of integration variables given in Eq.~(\ref{wvskz}) to Eqs.~(\ref{bra3d}) and (\ref{ket3d}), we obtain
\begin{eqnarray}
|\Psi \rangle & = & \sum_s \int |{\bf K},\omega,s\rangle \Psi({\bf K},\omega,s)\ {{\rm d}^2 K\ {\rm d}\omega\over c^2 k_z\ (2\pi)^3}  \label{bra2d} \\
\langle \Psi | & = & \sum_s \int \Psi^*({\bf K},\omega,s) \langle {\bf K},\omega,s |\ {{\rm d}^2 K\ {\rm d}\omega\over c^2 k_z\ (2\pi)^3} , \label{ket2d}
\end{eqnarray}
where 
\begin{equation}
k_z = \left( {\omega^2\over c^2} - k_x^2 - k_y^2 \right)^{1/2} ,
\label{kzdef}
\end{equation}
and the momentum space wave function is given by 
\begin{equation}
\Psi({\bf K},\omega,s) = \langle {\bf K},\omega,s |\Psi \rangle .
\label{binp2d}
\end{equation}
The inverse two-dimensional Fourier transform of the momentum space wave function gives the position space wave function on the transverse plane at $z=0$.

Using Eqs.~(\ref{bra2d}), (\ref{ket2d}) and the orthogonality condition
\begin{eqnarray}
\langle {\bf K}_1,\omega_1,r|{\bf K}_2,\omega_2,s\rangle & = & c^2 k_z\ (2\pi)^3 \delta_{r,s}\ \delta({\bf K}_1-{\bf K}_2) \nonumber \\ 
& & \times \delta(\omega_1-\omega_2) ,
\label{inprodk1}
\end{eqnarray}
one obtains the normalization condition of the one-photon states in terms of the two-dimensional momentum states
\begin{equation}
1 = \sum_s \int \left| \Psi({\bf K},\omega,s) \right|^2\ \ {{\rm d}^2 K\ {\rm d}\omega\over c^2 k_z\ (2\pi)^3} .
\label{norm1}
\end{equation}

\subsection{Monochromatic assumption}

The current application assumes that the optical field is monochromatic. For this purpose the momentum space wave function is assumed to be given by
\begin{equation}
\Psi({\bf K},\omega,s)= c\ G({\bf K}) H(\omega-\omega_0;\delta\omega) .
\label{wfdef}
\end{equation}
where $\omega_0$ and $\delta\omega$ are, respectively, the center frequency and the (small) bandwidth of the optical field. From now on we also drop the spin $s$, because we assume a uniform polarization. To satisfy the normalization requirement for the momentum space wave function given in Eq.~(\ref{norm1}) we define
\begin{equation}
H(\omega;\delta\omega)= \sqrt{2\sqrt{\pi}\over\delta\omega} \exp\left[-{\omega^2\over 2\delta\omega^2}\right] ,
\label{hdef}
\end{equation}
so that
\begin{equation}
\int \left| H(\omega;\delta\omega) \right|^2 {{\rm d} \omega\over 2\pi}  = 1 .
\label{normh}
\end{equation}
This definition of $H(\omega;\delta\omega)$ does not actually require the field to be monochromatic, unless we take $\delta\omega\rightarrow 0$. If we assume that the bandwidth $\delta\omega$ is very small, we can substitute $\omega=\omega_0$ inside $k_z$. As a result the normalization then reduces to
\begin{equation}
1 = \int \left| G({\bf K}) \right|^2 {{\rm d}^2 k\over 4\pi^2 k_z} .
\label{norm2}
\end{equation}
where
\begin{equation}
k_z = \left( {\omega_0^2\over c^2} - k_x^2 - k_y^2 \right)^{1/2} .
\label{kzdef0}
\end{equation}

The monochromatic two-dimensional momentum states can be defined by
\begin{eqnarray}
|{\bf K} \rangle & = & \int |{\bf K},\omega\rangle H(\omega-\omega_0;\delta\omega)\ {{\rm d} \omega\over 2\pi c}  \label{kbra} \\
\langle {\bf K} | & = & \int H(\omega-\omega_0;\delta\omega) \langle {\bf K},\omega |\ {{\rm d} \omega\over 2\pi c} , \label{kket}
\end{eqnarray}
and they obey the orthogonality relations
\begin{equation}
\langle {\bf K}_1|{\bf K}_2\rangle= 4\pi^2 k_z\ \delta({\bf K}_1-{\bf K}_2) .
\label{inprodk2}
\end{equation}

Using Eqs.~(\ref{kbra}) and (\ref{kket}) we can define the monochromatic one-photon states as follows
\begin{eqnarray}
|\Psi \rangle & = & \int |{\bf K}\rangle G({\bf K})\ {{\rm d}^2 K\over 4\pi^2 k_z }  \label{brak} \\
\langle \Psi | & = & \int G^*({\bf K}) \langle {\bf K} |\ {{\rm d}^2 K\over 4\pi^2 k_z} , \label{ketk}
\end{eqnarray}
in terms of the monochromatic two-dimensional momentum states. Note that we inserted $c$'s into the definitions of the monochromatic two-dimensional momentum states in Eqs.~(\ref{kbra}) and (\ref{kket}) and into Eq.~(\ref{wfdef}) so that the final definitions of the one-photon states in Eqs.~(\ref{brak}) and (\ref{ketk}) are without $c$'s.

\section{Density operator in the OAM basis}
\label{denop}

The set of all Laguerre-Gaussian (LG) modes forms a complete orthonormal basis. These modes are distinguished by two indices:\ an azimuthal index $\ell$ and a radial index $r$. For notational convenience these two indices are here combined into one index $m\equiv \{\ell,r\}$. Unless stated otherwise, each index used in the subsequent derivation always represents both the indices associated with a particular LG mode.

The LG modes are eigenstates of the rotations on the transverse plane. Since rotation invariance represents conservation of orbital angular momentum (OAM), these LG modes are also OAM eigenstates. In fact, the amount of OAM of an LG mode is proportional to the azimuthal index $\ell$ of that mode. The LG modes therefore form an OAM basis.

For the purpose of this derivation we first consider a single photon and then generalize the result for the case of two photons. The density operator of an arbitrary single photon state can be expressed in the OAM basis by
\begin{equation}
\rho = \sum_{m,n} |m\rangle\ \rho_{m,n}\  \langle n| .
\label{rhooam}
\end{equation}
Each of these OAM states can be expanded in terms of the monochromatic two-dimensional momentum basis, using Eqs.~(\ref{brak}) and (\ref{ketk}),
\begin{eqnarray}
|m \rangle & = & \int |{\bf K}\rangle G_m({\bf K})\ {{\rm d}^2 K\over 4\pi^2 k_z }  \label{bram} \\
\langle m | & = & \int G_m^*({\bf K}) \langle {\bf K} |\ {{\rm d}^2 K\over 4\pi^2 k_z} , \label{ketm}
\end{eqnarray}
leading to the following expression for the density operator in terms of the monochromatic two-dimensional momentum basis
\begin{eqnarray}
\rho(z) & = & \sum_{m,n} \int |{\bf K}_1\rangle\  G_m({\bf K}_1,z) \rho_{m,n} \nonumber \\ & & \times G_n^*({\bf K}_2,z) \langle{\bf K}_2|\ 
{{\rm d}^2 K_1\over 4\pi^2 k_{z1}} {{\rm d}^2 K_2\over 4\pi^2 k_{z2}} ,
\label{spk}
\end{eqnarray}
where the dependence on $z$ is shown explicitly to make it apparent that this expression for $\rho$ is only valid on a transverse plane for a specific value of $z$. Here, and also later, we use only one integral sign to represent several $K$-space integrals.

By evaluating the summations in Eq.~(\ref{spk}), one obtains the definition for the density operator in terms of the two-dimensional momentum basis,
\begin{equation}
\rho(z) = \int |{\bf K}_1\rangle \rho({\bf K}_1,{\bf K}_2,z) \langle {\bf K}_2|\ {{\rm d}^2 K_1\over 4\pi^2 k_{z1}} {{\rm d}^2 K_2\over 4\pi^2 k_{z2}} ,
\label{rhok}
\end{equation}
where
\begin{equation}
\rho({\bf K}_1,{\bf K}_2,z) = \sum_{m,n} G_m({\bf K}_1,z) \rho_{m,n} G_n^*({\bf K}_2,z) .
\label{oamk}
\end{equation}
The OAM eigenstates are orthonormal, $\langle m|n \rangle=\delta_{m,n}$, which, together with Eq.~(\ref{inprodk2}), implies that the momentum space wave functions $G_m({\bf K},z)$ are also orthonormal
\begin{equation}
\int G_m({\bf K},z) G_n^*({\bf K},z)\ {{\rm d}^2 K\over 4\pi^2 k_z} = \delta_{m,n} .
\label{ortowf}
\end{equation}
Using Eqs.~(\ref{oamk}) and (\ref{ortowf}) one can show that
\begin{eqnarray}
\rho_{m,n} & = & \int G_m^*({\bf K}_1,z) \rho({\bf K}_1,{\bf K}_2,z) \nonumber \\ & & \times G_n({\bf K}_2,z)\ {{\rm d}^2 K_1\over 4\pi^2 k_{z1}} {{\rm d}^2 K_2\over 4\pi^2 k_{z2}} .
\label{koam}
\end{eqnarray}

Since the two-dimensional momentum basis and the OAM basis are completely equivalent, the definitions in Eqs.~(\ref{rhooam}) and (\ref{rhok}) are also completely equivalent and Eqs.~(\ref{oamk}) and (\ref{koam}) indicate how one can transform from one to the other.

For two photons the density operator in Eq.~(\ref{spk}) can be generalized to become
\begin{eqnarray}
\rho(z) & = & \sum_{m,n,p,q} \int |{\bf K}_1\rangle |{\bf K}_3\rangle G_m({\bf K}_1,z) G_p({\bf K}_3,z) \nonumber \\ & & \times \rho_{m,n,p,q}\ G_n^*({\bf K}_2,z) G_q^*({\bf K}_4,z) \langle{\bf K}_2| \langle{\bf K}_4| \nonumber \\ & & \times {{\rm d}^2 K_1\over 4\pi^2 k_{z1}} {{\rm d}^2 K_2\over 4\pi^2 k_{z2}} {{\rm d}^2 K_3\over 4\pi^2 k_{z3}} {{\rm d}^2 K_4\over 4\pi^2 k_{z4}} .
\label{digt4}
\end{eqnarray}
For notational convenience we represent the product of momentum space wave functions that appear in Eq.~(\ref{digt4}), as a single function,
\begin{eqnarray}
& & G_m({\bf K}_1,z) G_n^*({\bf K}_2,z) G_p({\bf K}_3,z) G_q^*({\bf K}_4,z) \nonumber \\ & = & F_{m,n,p,q}({\bf K}_1,{\bf K}_2,{\bf K}_3,{\bf K}_4,z) .
\end{eqnarray}
The expression for $\rho$ is now given by
\begin{eqnarray}
\rho(z) & = & \sum_{m,n,p,q} \rho_{m,n,p,q} \int |{\bf K}_1\rangle |{\bf K}_3\rangle \nonumber \\ & & \times F_{m,n,p,q}({\bf K}_1,{\bf K}_2,{\bf K}_3,{\bf K}_4,z) \langle{\bf K}_2|\langle{\bf K}_4| \nonumber \\ & & \times {{\rm d}^2 K_1\over 4\pi^2 k_{z1}} {{\rm d}^2 K_2\over 4\pi^2 k_{z2}} {{\rm d}^2 K_3\over 4\pi^2 k_{z3}} {{\rm d}^2 K_4\over 4\pi^2 k_{z4}} .
\label{digt4s}
\end{eqnarray}

Note that inside the expression in Eq.~(\ref{digt4s}) the $z$-dependence is carried by the momentum space wave functions and not by the density matrix elements. This is because the transformation of the density operator during propagation over an infinitesimal distance through turbulence is caused by the distortion of the momentum space wave functions. After such an infinitesimal propagation these momentum space wave functions no longer represent the Fourier transforms of the original modes. One needs to re-expand these distorted wave functions in terms of the momentum space wave functions of the OAM modes and incorporate the expansion coefficients in the density matrix elements. Thereby one can transfer the $z$-dependence to the density matrix elements.

To obtain the full three-dimensional expression for the density operator in free-space (without turbulence) one can use Fresnel diffraction theory to determine the expression at any other value of $z$. In the presence of turbulence the expression for the density operator is only valid on a specific transverse plane, and it needs to be transformed from plane to plane, according to the dynamics of the medium [see Eq.~(\ref{spekprop}) below].

Since $F_{m,n,p,q}$ carries the only $z$-dependence in the expression for the density operator, we focus on how it transforms during infinitesimal propagation. At the end we apply the transformation to the expression for $\rho$.

\section{Equation of motion and infinitesimal transformation}
\label{eom}

For a classical electromagnetic field propagating through a source free region, the equation of motion, which follows directly from Maxwell's equations, is given by the Helmholtz equation,
\begin{equation}
\nabla^2 E({\bf x}) + n^2 k^2 E({\bf x}) = 0 ,
\end{equation}
where $n$ in the refractive index, $k$ is the wave number and ${\bf x} = x \hat{x} + y \hat{y} + z \hat{z}$. It is assumed that the polarization is uniform and can be ignored, therefore, only the scalar part of the electric field $E({\bf x})$ is considered here. The inhomogeneous medium is represented by a spatially varying index of refraction
\begin{equation}
n = 1 + \delta n ({\bf x}) .
\end{equation}
This variation is very small ($\delta n \ll 1$), which implies that one can approximate the Helmholtz equation as
\begin{equation}
\nabla^2 E({\bf x}) + k^2 E({\bf x}) + 2 \delta n({\bf x}) k^2 E({\bf x}) = 0 .
\end{equation}
Furthermore, we assume that the beam is paraxial and propagates in the $z$-direction. So we define
\begin{equation}
E({\bf x}) = g({\bf x}) \exp(-i k z) ,
\end{equation}
which then leads to the paraxial wave equation with an extra inhomogeneous medium term
\begin{equation}
\nabla_T^2 g({\bf x}) - i 2 k \partial_z g({\bf x}) + 2 \delta n({\bf x}) k^2 g({\bf x}) = 0 ,
\label{hhn}
\end{equation}
where $\nabla_T$ is the transverse part of the gradient operator.

The two-dimensional inverse Fourier transform,
\begin{equation}
g({\bf x}) = \int G({\bf K},z) \exp(-i {\bf K} \cdot {\bf x})\ {{\rm d}^2 K\over 4\pi^2} ,
\label{invft}
\end{equation}
which contains the angular spectrum of the optical field $G({\bf K},z)$, is substituted into Eq.~(\ref{hhn}), to obtain
\begin{equation}
\partial_z G({\bf K},z) = {i\over 2k} |{\bf K}|^2 G({\bf K},z) - i k N({\bf K},z) \star G({\bf K},z) ,
\label{spekprop}
\end{equation}
where $\star$ indicates convolution and $N({\bf K},z)$ is the two-dimensional Fourier transform of $\delta n({\bf x})$. The $z$-dependence is shown explicitly to indicate that $G({\bf K},z)$ depends on the location of the transverse plane along $z$.

Note that the (inverse) Fourier transform in Eq.~(\ref{invft}) is a purely formal operation on the two-dimensional function and therefore does not contain a factor of $1/k_z$. The convolution integral, which comes from the Fourier transform of the product of two functions, does not contain a factor of $1/k_z$ either.

\begin{figure}[ht]
\centerline{\scalebox{1}{\includegraphics{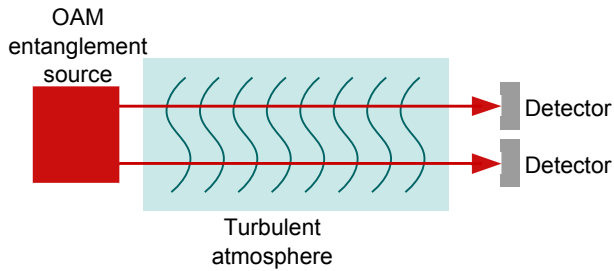}}}
\caption{Diagram of an OAM entangled biphoton propagating through a turbulent medium toward two detectors.}
\label{bron}
\end{figure}

The quantum wave function in an interaction-free system obeys the same equation of motion as the classical field. As a result, one can identify the momentum space quantum wave function with the angular spectrum in Eq.~(\ref{spekprop}). Thus the expression in Eq.~(\ref{spekprop}) represents the infinitesimal transformation of the momentum space wave function during propagation through a turbulent (random) medium. It forms the basis of the derivation of the IPE.

\section{Derivation of the IPE}
\label{deripe}

Here we consider the scenario where two photons (a biphoton), which are entangled in terms of the OAM basis, both propagate through a turbulent atmosphere, as shown in Fig.~\ref{bron}. The derivation of the IPE follows the same procedure that was followed in \cite{ipe}. The only differences are that the integration measure for the phase space integrals contain an extra $1/k_z$-factor and here we allow the two photons to go through the same turbulent medium, which means that there are additional correlations that were excluded in \cite{ipe}.

\subsection{Infinitesimal transformation}

First we consider what happens to $F_{m,n,p,q}$ after an infinitesimal propagation, by setting $z\rightarrow z+dz$. Expanding the result to sub-leading order in $dz$, applying Eq.~(\ref{spekprop}) and taking the limit $dz\rightarrow 0$, we obtain a differential equation given by
\begin{eqnarray}
& & \partial_z F_{m,n,p,q}({\bf K}_1,{\bf K}_2,{\bf K}_3,{\bf K}_4,z) \nonumber \\ 
& = & {i\over 2k} (|{\bf K}_1|^2-|{\bf K}_2|^2+|{\bf K}_3|^2-|{\bf K}_4|^2) \nonumber \\
& & \times F_{m,n,p,q}({\bf K}_1,{\bf K}_2,{\bf K}_3,{\bf K}_4,z) \nonumber \\
& & - i k \int N({\bf K}_1-{\bf K},z) F_{m,n,p,q}({\bf K},{\bf K}_2,{\bf K}_3,{\bf K}_4,z) \nonumber \\
& & - N^*({\bf K}_2-{\bf K},z) F_{m,n,p,q}({\bf K}_1,{\bf K},{\bf K}_3,{\bf K}_4,z) \nonumber \\
& & + N({\bf K}_3-{\bf K},z) F_{m,n,p,q}({\bf K}_1,{\bf K}_2,{\bf K},{\bf K}_4,z) \nonumber \\
& & - N^*({\bf K}_4-{\bf K},z) F_{m,n,p,q}({\bf K}_1,{\bf K}_2,{\bf K}_3,{\bf K},z) \nonumber \\
& & \times {{\rm d}^2 K\over 4\pi^2} .
\label{ensf0}
\end{eqnarray}
The integral in Eq.~(\ref{ensf0}) does not contain the additional $1/k_z$-factor because it is a convolution integral.

Since the spectra $N({\bf K},z)$ are random functions, $F_{m,n,p,q}$ in Eq.~(\ref{ensf0}) is also a random function. To calculate the expectation value for $F_{m,n,p,q}$ we need to employ ensemble averaging. (In a slight abuse of notation we'll keep on denoting the ensemble average of $F_{m,n,p,q}$ simply as $F_{m,n,p,q}$.) For this purpose we start by integrating Eq.~(\ref{ensf0}) from $z_0$ to $z$, which gives $F_{m,n,p,q}$ in terms of previous versions of itself. Using repeated back substitution, one obtains a Born series (or Dyson expansion). Since each $N({\bf K},z)$ effectively comes with a factor of $C_n$ (the square root of the refractive index structure constant $C_n^2$), higher orders in $N({\bf K},z)$ are suppressed. However, since $\langle N({\bf K},z) \rangle=0$, one needs to expand the series at least up to second order in $N({\bf K},z)$. The resulting expression, without the $\langle N({\bf K},z) \rangle$-terms, is given by
\begin{widetext}
\begin{eqnarray}
F_{m,n,p,q}({\bf K}_1,{\bf K}_2,{\bf K}_3,{\bf K}_4,z) & = & F_{m,n,p,q}({\bf K}_1,{\bf K}_2,{\bf K}_3,{\bf K}_4,z_0) \nonumber \\
& & + (z-z_0) {i\over 2k} (|{\bf K}_1|^2-|{\bf K}_2|^2+|{\bf K}_3|^2-|{\bf K}_4|^2) F_{m,n,p,q}({\bf K}_1,{\bf K}_2,{\bf K}_3,{\bf K}_4,z_0) \nonumber \\
& & - k^2\int_{z_0}^{z} \int_{z_0}^{z_1} \int 
\langle N({\bf K}_1-{\bf K},z_1) N({\bf K}-{\bf K}_0,z_2) \rangle F_{m,n,p,q}({\bf K}_0,{\bf K}_2,{\bf K}_3,{\bf K}_4,z_0) \nonumber \\
& & - \langle N({\bf K}_1-{\bf K},z_1) N^*({\bf K}_2-{\bf K}_0,z_2) \rangle F_{m,n,p,q}({\bf K},{\bf K}_0,{\bf K}_3,{\bf K}_4,z_0) \nonumber \\
& & + \langle N({\bf K}_1-{\bf K},z_1) N({\bf K}_3-{\bf K}_0,z_2) \rangle F_{m,n,p,q}({\bf K},{\bf K}_2,{\bf K}_0,{\bf K}_4,z_0) \nonumber \\
& & - \langle N({\bf K}_1-{\bf K},z_1) N^*({\bf K}_4-{\bf K}_0,z_2) \rangle F_{m,n,p,q}({\bf K},{\bf K}_2,{\bf K}_3,{\bf K}_0,z_0) \nonumber \\
& & - \langle N^*({\bf K}_2-{\bf K},z_1) N({\bf K}_1-{\bf K}_0,z_2) \rangle F_{m,n,p,q}({\bf K}_0,{\bf K},{\bf K}_3,{\bf K}_4,z_0) \nonumber \\
& & + \langle N^*({\bf K}_2-{\bf K},z_1) N^*({\bf K}-{\bf K}_0,z_2) \rangle F_{m,n,p,q}({\bf K}_1,{\bf K}_0,{\bf K}_3,{\bf K}_4,z_0) \nonumber \\
& & - \langle N^*({\bf K}_2-{\bf K},z_1) N({\bf K}_3-{\bf K}_0,z_2) \rangle F_{m,n,p,q}({\bf K}_1,{\bf K},{\bf K}_0,{\bf K}_4,z_0) \nonumber \\
& & + \langle N^*({\bf K}_2-{\bf K},z_1) N^*({\bf K}_4-{\bf K}_0,z_2) \rangle F_{m,n,p,q}({\bf K}_1,{\bf K},{\bf K}_3,{\bf K}_0,z_0) \nonumber \\
& & + \langle N({\bf K}_3-{\bf K},z_1) N({\bf K}_1-{\bf K}_0,z_2) \rangle F_{m,n,p,q}({\bf K}_0,{\bf K}_2,{\bf K},{\bf K}_4,z_0) \nonumber \\
& & - \langle N({\bf K}_3-{\bf K},z_1) N^*({\bf K}_2-{\bf K}_0,z_2) \rangle F_{m,n,p,q}({\bf K}_1,{\bf K}_0,{\bf K},{\bf K}_4,z_0) \nonumber \\
& & + \langle N({\bf K}_3-{\bf K},z_1) N({\bf K}-{\bf K}_0,z_2) \rangle F_{m,n,p,q}({\bf K}_1,{\bf K}_2,{\bf K}_0,{\bf K}_4,z_0) \nonumber \\
& & - \langle N({\bf K}_3-{\bf K},z_1) N^*({\bf K}_4-{\bf K}_0,z_2) \rangle F_{m,n,p,q}({\bf K}_1,{\bf K}_2,{\bf K},{\bf K}_0,z_0) \nonumber \\
& & - \langle N^*({\bf K}_4-{\bf K},z_1) N({\bf K}_1-{\bf K}_0,z_2) \rangle F_{m,n,p,q}({\bf K}_0,{\bf K}_2,{\bf K}_3,{\bf K},z_0) \nonumber \\
& & + \langle N^*({\bf K}_4-{\bf K},z_1) N^*({\bf K}_2-{\bf K}_0,z_2) \rangle F_{m,n,p,q}({\bf K}_1,{\bf K}_0,{\bf K}_3,{\bf K},z_0) \nonumber \\
& & - \langle N^*({\bf K}_4-{\bf K},z_1) N({\bf K}_3-{\bf K}_0,z_2) \rangle F_{m,n,p,q}({\bf K}_1,{\bf K}_2,{\bf K}_0,{\bf K},z_0) \nonumber \\
& & + \langle N^*({\bf K}_4-{\bf K},z_1) N^*({\bf K}-{\bf K}_0,z_2) \rangle F_{m,n,p,q}({\bf K}_1,{\bf K}_2,{\bf K}_3,{\bf K}_0,z_0) \nonumber \\
& & \times {{\rm d}^2 K_0\over 4\pi^2}\ {{\rm d}^2 K\over 4\pi^2}\ {\rm d} z_2\ {\rm d} z_1  .
\label{ensf}
\end{eqnarray}
In Appendix \ref{ensemb} it is shown that
\begin{equation}
\int_{z_0}^{z} \int_{z_0}^{z_1} \langle N({\bf K}_1,z_2) N^*({\bf K}_2,z_1) \rangle\ {\rm d} z_2\ {\rm d} z_1 = 2\pi^2 dz \delta({\bf K}_1-{\bf K}_2) \Phi_1({\bf K}_1) ,
\label{verw3}
\end{equation}
where we set $z-z_0=dz$. 

We now use Eq.~(\ref{verw3}) to simplify Eq.~(\ref{ensf})  and then take the limit $dz\rightarrow 0$, to turn it into a differential equation again,
\begin{eqnarray}
\partial_z F_{m,n,p,q}({\bf K}_1,{\bf K}_2,{\bf K}_3,{\bf K}_4,z) & = &  {i\over 2k} (|{\bf K}_1|^2-|{\bf K}_2|^2+|{\bf K}_3|^2-|{\bf K}_4|^2) F_{m,n,p,q}({\bf K}_1,{\bf K}_2,{\bf K}_3,{\bf K}_4,z) \nonumber \\
& & - 2 k^2 F_{m,n,p,q}({\bf K}_1,{\bf K}_2,{\bf K}_3,{\bf K}_4,z) \int \Phi_1({\bf K})\ {{\rm d}^2 K\over 4\pi^2} \nonumber \\
& & + k^2 \int  \Phi_1({\bf K}) F_{m,n,p,q}({\bf K}_1-{\bf K},{\bf K}_2-{\bf K},{\bf K}_3,{\bf K}_4,z)\ {{\rm d}^2 K\over 4\pi^2} \nonumber \\
& & + k^2 \int  \Phi_1({\bf K}) F_{m,n,p,q}({\bf K}_1,{\bf K}_2-{\bf K},{\bf K}_3-{\bf K},{\bf K}_4,z)\ {{\rm d}^2 K\over 4\pi^2} \nonumber \\
& & + k^2 \int  \Phi_1({\bf K}) F_{m,n,p,q}({\bf K}_1-{\bf K},{\bf K}_2,{\bf K}_3,{\bf K}_4-{\bf K},z)\ {{\rm d}^2 K\over 4\pi^2} \nonumber \\
& & + k^2 \int  \Phi_1({\bf K}) F_{m,n,p,q}({\bf K}_1,{\bf K}_2,{\bf K}_3-{\bf K},{\bf K}_4-{\bf K},z)\ {{\rm d}^2 K\over 4\pi^2} \nonumber \\
& & - k^2 \int \Phi_1({\bf K}) F_{m,n,p,q}({\bf K}+{\bf K}_1,{\bf K}_2,{\bf K}_3-{\bf K},{\bf K}_4,z)\ {{\rm d}^2 K\over 4\pi^2} \nonumber \\
& & - k^2 \int \Phi_1({\bf K}) F_{m,n,p,q}({\bf K}_1,{\bf K}+{\bf K}_2,{\bf K}_3,{\bf K}_4-{\bf K},z)\ {{\rm d}^2 K\over 4\pi^2} .
\label{ensf1}
\end{eqnarray}
If one substitutes Eq.~(\ref{ensf1}) into the $z$-derivative of Eq.~(\ref{digt4s}), one would obtain a first order differential equation for the density operator. However, we are interested in the transformation of the individual density matrix elements.

\subsection{Extraction of matrix elements}

To express the transformation of the density operator due to the infinitesimal propagation through a turbulent atmosphere in terms of the density matrix elements, we extract the matrix elements from the density operator using the trace
\begin{equation}
\partial_z \rho_{u,v,r,s}(z) = {\rm trace} \left\{ \partial_z \rho(z) |v\rangle |s\rangle \langle u| \langle r| \right\} ,
\label{trace0}
\end{equation}
where the operator that selects a particular matrix element in the OAM basis is given by
\begin{equation}
|v\rangle |s\rangle \langle u| \langle r| = \frac{1}{4} \int |{\bf K}_8\rangle |{\bf K}_6\rangle\ F_{u,v,r,s}^*({\bf K}_5,{\bf K}_6,{\bf K}_7,{\bf K}_8,z)
\langle{\bf K}_7|\langle{\bf K}_5|\ {{\rm d}^2 K_5\over 4\pi^2 k_{z5}} {{\rm d}^2 K_6\over 4\pi^2 k_{z6}} {{\rm d}^2 K_7\over 4\pi^2 k_{z7}} {{\rm d}^2 K_8\over 4\pi^2 k_{z8}} .
\label{digt5}
\end{equation}
The factor of a quarter in Eq.~(\ref{digt5}) comes from the fact that, propagating through the same medium, the two photons are indistinguishable. As a result there are four different ways in which the states can be contracted on each other. This gives rise to a symmetry factor of 4, which implies that the same terms are counted several times. To remove this over counting, one needs to insert the factor of a quarter in Eq.~(\ref{digt5}). 

We now substitute the derivative of Eqs.~(\ref{digt4s}) with respect to $z$ and (\ref{digt5}) into Eq.~(\ref{trace0}) to obtain
\begin{eqnarray}
\partial_z \rho_{u,v,r,s}(z) & = &  {\rm trace} \left\{ \sum_{m,n,p,q} \rho_{m,n,p,q} \int |{\bf K}_1\rangle |{\bf K}_3\rangle \partial_z F_{m,n,p,q}({\bf K}_1,{\bf K}_2,{\bf K}_3,{\bf K}_4,z) \right. \nonumber \\
& & \times \langle{\bf K}_2|\langle{\bf K}_4|\ {{\rm d}^2 K_1\over 4\pi^2 k_{z1}} {{\rm d}^2 K_2\over 4\pi^2 k_{z2}} {{\rm d}^2 K_3\over 4\pi^2 k_{z3}} {{\rm d}^2 K_4\over 4\pi^2 k_{z4}} \nonumber \\
& & \left. \times \frac{1}{4} \int |{\bf K}_8\rangle |{\bf K}_6\rangle\ F_{u,v,r,s}^*({\bf K}_5,{\bf K}_6,{\bf K}_7,{\bf K}_8,z) \langle{\bf K}_7|\langle{\bf K}_5|
\ {{\rm d}^2 K_5\over 4\pi^2 k_{z5}} {{\rm d}^2 K_6\over 4\pi^2 k_{z6}} {{\rm d}^2 K_7\over 4\pi^2 k_{z7}} {{\rm d}^2 K_8\over 4\pi^2 k_{z8}} \right \} \nonumber \\
& = & \sum_{m,n,p,q} \rho_{m,n,p,q} \int \partial_z F_{m,n,p,q}({\bf K}_1,{\bf K}_2,{\bf K}_3,{\bf K}_4,z) \nonumber \\
& & \times F_{u,v,r,s}^*({\bf K}_1,{\bf K}_2,{\bf K}_3,{\bf K}_4,z)\ {{\rm d}^2 K_1\over 4\pi^2 k_{z1}} {{\rm d}^2 K_2\over 4\pi^2 k_{z2}} {{\rm d}^2 K_3\over 4\pi^2 k_{z3}} {{\rm d}^2 K_4\over 4\pi^2 k_{z4}} ,
\label{digt6}
\end{eqnarray}
where the last expression is obtained because of the orthogonality of the momentum basis, given in Eq.~(\ref{inprodk2}). The factor of 4 that comes from the multiple ways in which the momentum states can be contracted on each other removed the factor of a quarter in the last expression. Substituting Eq.~(\ref{ensf1}) into Eq.~(\ref{digt6}), we obtain
\begin{eqnarray}
\partial_z \rho_{u,v,r,s}(z) & = &  \sum_{m,n,p,q} \rho_{m,n,p,q} \int \left[ {i \over 2k} (|{\bf K}_1|^2-|{\bf K}_2|^2+|{\bf K}_3|^2-|{\bf K}_4|^2) F_{m,n,p,q}({\bf K}_1,{\bf K}_2,{\bf K}_3,{\bf K}_4,z) \right. \nonumber \\
& & - 2 k^2 F_{m,n,p,q}({\bf K}_1,{\bf K}_2,{\bf K}_3,{\bf K}_4,z) \int \Phi_1({\bf K})\ {{\rm d}^2 K\over 4\pi^2} \nonumber \\
& & + k^2 \int \Phi_1({\bf K}) F_{m,n,p,q}({\bf K}_1-{\bf K},{\bf K}_2-{\bf K},{\bf K}_3,{\bf K}_4,z)\ {{\rm d}^2 K\over 4\pi^2} \nonumber \\
& & + k^2 \int \Phi_1({\bf K}) F_{m,n,p,q}({\bf K}_1,{\bf K}_2-{\bf K},{\bf K}_3-{\bf K},{\bf K}_4,z)\ {{\rm d}^2 K\over 4\pi^2} \nonumber \\
& & + k^2 \int \Phi_1({\bf K}) F_{m,n,p,q}({\bf K}_1-{\bf K},{\bf K}_2,{\bf K}_3,{\bf K}_4-{\bf K},z)\ {{\rm d}^2 K\over 4\pi^2} \nonumber \\
& & + k^2 \int \Phi_1({\bf K}) F_{m,n,p,q}({\bf K}_1,{\bf K}_2,{\bf K}_3-{\bf K},{\bf K}_4-{\bf K},z)\ {{\rm d}^2 K\over 4\pi^2} \nonumber \\
& & - k^2 \int \Phi_1({\bf K}) F_{m,n,p,q}({\bf K}+{\bf K}_1,{\bf K}_2,{\bf K}_3-{\bf K},{\bf K}_4,z)\ {{\rm d}^2 K\over 4\pi^2} \nonumber \\
& & \left. - k^2 \int \Phi_1({\bf K}) F_{m,n,p,q}({\bf K}_1,{\bf K}+{\bf K}_2,{\bf K}_3,{\bf K}_4-{\bf K},z)\ {{\rm d}^2 K\over 4\pi^2} \right] \nonumber \\ 
& & \times F_{u,v,r,s}^*({\bf K}_1,{\bf K}_2,{\bf K}_3,{\bf K}_4,z)\ {{\rm d}^2 K_1\over 4\pi^2 k_{z1}} {{\rm d}^2 K_2\over 4\pi^2 k_{z2}} {{\rm d}^2 K_3\over 4\pi^2 k_{z3}} {{\rm d}^2 K_4\over 4\pi^2 k_{z4}} .
\label{ensf2}
\end{eqnarray}
The expressions in Eq.~(\ref{ensf2}) can be further simplified by using the orthogonality of the momentum space wave functions of the OAM basis given in Eq.~(\ref{ortowf}).
\end{widetext}

\subsection{Final expression}

After the final simplification of Eq.~(\ref{ensf2}) we obtain the IPE as a set of first order differential equations given by
\begin{eqnarray}
\partial_z \rho_{u,v,r,s} & = & S_{m,u} \rho_{m,v,r,s} - S_{v,n} \rho_{u,n,r,s} \nonumber \\
& & + S_{p,r} \rho_{u,v,p,s} - S_{s,q} \rho_{u,v,r,q} \nonumber \\
& & + L_{v,u,n,m} \rho_{m,n,r,s} + L_{v,r,n,p} \rho_{u,n,p,s} \nonumber \\
& & + L_{s,u,q,m} \rho_{m,v,r,q} + L_{s,r,q,p} \rho_{u,v,p,q} \nonumber \\
& & - L_{m,r,u,p} \rho_{m,v,p,s} - L_{s,n,q,v} \rho_{u,n,r,q} \nonumber \\
& & - 2 \rho_{u,v,r,s} L_T ,
\label{ensf3}
\end{eqnarray}
where repeated indices imply summation. The quantities in Eq.~(\ref{ensf3}) are defined as follows
\begin{equation}
S_{x,y}(z) = {i \over 2k} \int |{\bf K}|^2 G_x({\bf K},z) G_y^*({\bf K},z)\ {{\rm d}^2 K\over 4\pi^2 k_z} ,
\label{defs}
\end{equation}
\begin{equation}
L_T = k^2 \int \Phi_1({\bf K})\ {{\rm d}^2 K\over 4\pi^2} ,
\label{deflt}
\end{equation}
and
\begin{equation}
L_{m,n,u,v}(z) = k^2 \int \Phi_1({\bf K}) W_{m,u}({\bf K},z) W_{n,v}^*({\bf K},z)\ {{\rm d}^2 K\over 4\pi^2} ,
\label{defl}
\end{equation}
with
\begin{equation}
W_{x,y}({\bf K},z) = \int G_x({\bf K}_1,z) G_y^*({\bf K}_1-{\bf K},z)\ {{\rm d}^2 K_1\over 4\pi^2 k_{z1}} .
\label{defw}
\end{equation}

The first four terms of Eq.~(\ref{ensf3}) are the non-dissipative terms, representing the free-space propagation process. The last seven terms of Eq.~(\ref{ensf3}) are the dissipative terms, representing the scattering of OAM modes into other OAM modes due to scintillation caused by the turbulence.

\section{Solving the integrals}
\label{solvint}

The solution of the integrals in Eqs.~(\ref{defs})-(\ref{defw}) have been considered in \cite{ipe}. Two of these expressions, Eqs.~(\ref{defs}) and (\ref{defw}), are now slightly different due to the presence of the $1/k_z$-factor. Moreover, to satisfy the orthogonality condition for the momentum space wave functions $G_m({\bf K},z)$, the generating function for the LG modes that was proposed in \cite{ipe}, now needs to contain an additional factor of the square root of $k_z$. The expression for the generating function of the spectra of the LG modes is therefore given by 
\begin{eqnarray}
{\cal F} \{G\} & = & \frac{\pi k_z^{1/2}}{1+w}\exp \left[ { i\pi (a+ib)p \over 1+w } + {i\pi (a-ib)q \over 1+w } \right. \nonumber \\ 
& & \left. - {\pi^2 (a^2+b^2)(1-w-it-iwt) \over 1+w } \right]
\label{msgf1}
\end{eqnarray}
where $p$, $q$, and $w$ are used to generated the spectrum of a particular LG mode, $a$ and $b$ are normalized spatial frequency components related to $k_x$ and $k_y$ via
\begin{equation}
k_x = {2\pi a\over d_0} ~~~~~~~~~ k_y = {2\pi b\over d_0} ,
\end{equation}
and
\begin{equation}
k_z^{1/2} = \left[ {\omega_0^2\over c^2} - {4\pi^2\over d_0^2} \left(a^2+b^2\right) \right]^{1/4} .
\end{equation}
with $\omega_0$ and $d_0$ being the centre frequency and beam radius, respectively.

When the generating function in Eq.~(\ref{msgf1}) is substituted into Eq.~(\ref{defs}) the $k_z$-factors cancel, leaving the same expression that was evaluated in \cite{ipe}.

The integral in Eq.~(\ref{defw}) represents the correlation between the momentum space wave functions of the OAM modes. In this case the $k_z$-factors do not cancel, but leave a factor of
\begin{equation}
\left({k_z'\over k_z}\right)^{1/2} = \left[ { d_0^2 - \lambda^2 (a_1-a)^2 - \lambda^2 (b_1-b)^2 \over d_0^2 - \lambda^2 a_1^2 - \lambda^2 b_1^2} \right]^{1/4} ,
\end{equation}
where $\lambda$ is the wavelength associated with $\omega_0$. In the paraxial limit ($\lambda\ll d_0$), this factor becomes
\begin{equation}
\left({k_z'\over k_z}\right)^{1/2} \sim 1 + O \left( \frac{\lambda^2}{d_0^2} \right) .
\end{equation}
Therefore, one can neglect this factor, rendering the resulting expression in the same form as that which was evaluated in \cite{ipe}.

Hence, in the paraxial limit all the integrals in Eqs.~(\ref{defs})-(\ref{defw}) are the same as those that were considered in \cite{ipe}. The current formalism that starts from Lorentz invariant definitions of the quantum states does not produce expressions that are significantly different from those that were obtained in \cite{ipe}. 

\section{Conclusions}

An expression is obtained for the evolution of the density matrix elements of a spatial mode entangled bi-photon propagating through the same turbulent atmosphere. The expression represents an infinite set of coupled first order differential equations, each containing non-dissipative terms associated with free-space propagation and dissipative terms associated with modal scattering due to the scintillation. Among the dissipative terms are included terms associated with the cross correlation between the two photons, coming from the fact that they are indistinguishable photons propagating through the same medium.

The derivation starts from a manifestly Lorentz covariant formulation of the quantum states and then follows a number of steps that eventually explicitly break the Lorentz invariance. This leads to a final expression that differs from the previous expression in \cite{ipe} in that some of the integrals (the phase space integrals) in the new expression contain factors of $1/k_z$. The only place where these factors could potentially have an effect is in the correlation between different momentum space wave functions for the different OAM modes. However, in this case it is shown that the $1/k_z$-factors can be ignored in the paraxial limit. 

Hence, although the formal expression that is obtained from an initially Lorentz invariant formulation differs from what is obtained from the more traditional formulation, the effect of this difference on measurements is not expected to be significant or even observable.

\appendix

\section{Ensemble average}
\label{ensemb}

The evaluation of the ensemble averages of products of the random functions $N({\bf K},z)$ in Eq.~(\ref{ensf}), is discussed in an appendix in \cite{ipe}. For the sake of convenience this appendix is included here, with only minor changes.

As mentioned in Sec.~\ref{eom}, the refractive index fluctuation produced by a turbulent atmosphere is small compared to the average refractive index of air, $\delta n \ll 1$, which leads to the fact that one can separate the propagation through a turbulent atmosphere into two parts: free-space propagation and random phase modulations. The random phase functions for the latter step are obtained by integrating the refractive index fluctuation through a thin sheet of atmosphere along the propagation direction,
\begin{eqnarray}
\theta(x,y) & = & k \int_{z_0-\delta z/2}^{z_0+\delta z/2} \delta n(x,y,z) {\rm d} z \nonumber \\
& \approx & k\ \delta z\ \delta n(x,y,z_0) ,
\end{eqnarray}
where, in the last line we took the limit $\delta z \rightarrow 0$. Replacing the refractive index fluctuation with its Fourier expansion, we obtain
\begin{eqnarray}
\theta(x,y,z_0) & = & k \delta z \int \exp[-i(k_x x + k_y y + k_z z_0)] \nonumber \\
& & \times N({\bf k})\ {{\rm d}^3 k\over (2\pi)^3} ,
\end{eqnarray}
where $N({\bf k})$ is the three-dimensional spatial spectrum of the index fluctuation. We now define a two-dimensional spectrum for the accumulated index fluctuation $N({\bf K},z)$ over a thin sheet of atmosphere
\begin{equation}
N({\bf K},z) = \int \exp(-i k_z z) N({\bf k})\ {{\rm d} k_z\over 2\pi} ,
\label{spek2d}
\end{equation}
which depends on the $z$ position of the thin sheet. The three-dimensional spectrum of the refractive index fluctuation can be expressed in terms of its three-dimensional power spectral density, which follows from the autocorrelation function of the index fluctuation and which represents the model for the turbulence,
\begin{equation}
N({\bf k}) = \tilde{\chi}({\bf k}) \left[ {\Phi_0({\bf k})\over\Delta_k^3} \right]^{1/2} ,
\label{nspek}
\end{equation}
where $\tilde{\chi}({\bf k})$ is a normally distributed random complex spectral function and $\Delta_k$ is its spatial coherence length in the frequency domain. The latter is inversely proportional to the outer scale of the turbulence. Since the refractive index fluctuation $\delta n$ is an asymmetric real-valued function, we have that $\tilde{\chi}^*({\bf k})=\tilde{\chi}(-{\bf k})$. Furthermore, the autocorrelation function of the random function is
\begin{equation}
\langle \tilde{\chi}({\bf k}_1) \tilde{\chi}^*({\bf k}_2) \rangle = \left( 2\pi\Delta_k \right)^3 \delta_3 ({\bf k}_1-{\bf k}_2) .
\label{verwrand}
\end{equation}

In Eq.~(\ref{ensf}) we find ensemble averages inside double $z$-integrals. Substituting Eqs.~(\ref{spek2d}) and (\ref{nspek}) into such an ensemble average and using Eq.~(\ref{verwrand}) to evaluate the ensemble average, one obtains
\begin{eqnarray}
& & \int_{z_0}^{z} \int_{z_0}^{z_1} \langle N({\bf K}_1,z_2) N^*({\bf K}_2,z_1) \rangle\ {\rm d} z_2\ {\rm d} z_1 \nonumber \\
& = & (2\pi)^2 \delta({\bf K}_1-{\bf K}_2) \int \int_{z_0}^{z} \int_{z_0}^{z_1} \Phi_0({\bf k}_1) \nonumber \\
& & \times \exp \left[i k_z (z_1-z_2) \right]\ {\rm d} z_2\ {\rm d} z_1\ {{\rm d} k_z\over 2\pi}
\label{verw1}
\end{eqnarray}
We set $z=z_0+dz$ and evaluate the two $z$-integrals
\begin{eqnarray}
& & \int_{z_0}^{z_0+dz} \int_{z_0}^{z_1} \exp \left[i k_z (z_1-z_2) \right]\ {\rm d} z_2\ {\rm d} z_1 \nonumber \\
& = & {1-\cos(k_z dz)\over k_z^2} + i {\sin(k_z dz)-k_z dz\over  k_z^2} .
\label{zint}
\end{eqnarray}
The power spectral density $\Phi_0({\bf k}_1)$ is always even in $k_z$. Therefore, the imaginary part of Eq.~(\ref{zint}), being odd in $k_z$, does not contribute to the final expression,
\begin{eqnarray}
& & \int_{z_0}^{z} \int_{z_0}^{z_1} \langle N({\bf K}_1,z_2) N^*({\bf K}_2,z_1) \rangle\ {\rm d} z_2\ {\rm d} z_1 \nonumber \\
& = & (2\pi)^2 \delta({\bf K}_1-{\bf K}_2) \int \Phi_0({\bf k}_1) \nonumber \\ & & \times \left[ {1-\cos(k_z dz)\over k_z^2} \right] \ {{\rm d} k_z\over 2\pi} .
\label{verw2}
\end{eqnarray}
Due to the fact that the refractive index variations are very small, the light that propagates through the turbulent atmosphere remains unchanged over distances much longer than the correlation distance of the turbulent medium. One can therefore assume that $dz$ is much larger than this correlation distance. As a result the function inside the square-brackets in Eq.~(\ref{verw2}) acts like a Dirac delta function, so that one can substitute $k_z=0$ in $\Phi_0$ and pull it out of the $k_z$-integral. The integral can then be evaluated to give
\begin{eqnarray}
& & \int_{z_0}^{z_1} \int_{z_0}^{z} \langle N({\bf K}_1,z_2) N^*({\bf K}_2,z) \rangle\ {\rm d} z_2\ {\rm d} z \nonumber \\
& = & 2\pi^2 dz \delta({\bf K}_1-{\bf K}_2) \Phi_1({\bf K}_1) ,
\end{eqnarray}
where we defined $\Phi_1({\bf K}_1)=\Phi_0({\bf K}_1,0)$.

\end{document}